\documentclass[aps,prl,psfig,amsmath,epsfig,twocolumn]{revtex4}

\usepackage{graphics}
\input epsf

\def\be{\begin{equation}}
\def\ee{\end{equation}}
\def\bea{\begin{eqnarray}}
\def\eea{\end{eqnarray}}

\begin{document}

\title{\begin{boldmath} Measurements of R = $\sigma_L / \sigma_T$ and the 
Separated Longitudinal and Transverse Structure Functions in the Nucleon 
Resonance Region\end{boldmath}}

\author{
Y.~Liang,$^{1,7}$
V.~Tvaskis,$^{7,19}$
M.E.~Christy,$^{7}$
A.~Ahmidouch,$^{14}$ 
C.~S.~Armstrong,$^{19}$ 
J.~Arrington,$^{2}$
R.~Asaturyan,$^{23}$
S.~Avery,$^{7}$
O.~K.~Baker,$^{7,19}$
D.~H.~Beck,$^{8}$
H.~P.~Blok,$^{21}$
C.~W.~Bochna,$^{8}$
W.~Boeglin,$^{4,19}$
P.~Bosted,$^{1,19}$
M.~Bouwhuis,$^{8}$
H.~Breuer,$^{9}$
D.~S.~Brown,$^{9}$
A.~Bruell,$^{10}$
R.~D.~Carlini,$^{19}$
J.~Cha,$^{12}$
N.~S.~Chant,$^{9}$
A.~Cochran,$^{7}$
L.~Cole,$^{7}$
S.~Danagoulian,$^{14}$ 
D.~B.~Day,$^{20}$
J.~Dunne,$^{12}$
D.~Dutta,$^{10,12}$ 
R.~Ent,$^{19}$
H.~C.~Fenker,$^{19}$
B.~Fox,$^{3}$
L.~Gan,$^{7}$
H.~Gao,$^{10}$
K.~Garrow,$^{19}$
D.~Gaskell,$^{2,16}$
A.~Gasparian,$^{7}$
D.~F.~Geesaman,$^{2}$
R.~Gilman,$^{18,19}$
P.~L.~J.~Gu\`eye,$^{7}$
M.~Harvey,$^{7}$
R.~J.~Holt,$^{8}$
X.~Jiang,$^{18}$
M.~Jones,$^{19}$
C.~E.~Keppel,$^{7,19}$ 
E.~Kinney,$^{3}$
W.~Lorenzon,$^{11}$ 
A.~Lung,$^{19}$ 
D.~J.~Mack,$^{19}$
P.~Markowitz,$^{4,19}$
J.~W.~Martin,$^{10}$
K.~McIlhany,$^{10}$
D.~McKee,$^{13}$
D.~Meekins,$^{5,19}$ 
M.~A.~Miller,$^{8}$ 
R.~G.~Milner,$^{10}$ 
J.~H.~Mitchell,$^{19}$
H.~Mkrtchyan,$^{23}$
B.~A.~Mueller,$^{2}$
A.~Nathan,$^{8}$
G.~Niculescu,$^{15}$ 
I.~Niculescu,$^{6}$ 
T.~G.~O'Neill,$^{2}$
V.~Papavassiliou,$^{13,19}$
S.F.~Pate,$^{13,19}$
R.~B.~Piercey,$^{12}$ 
D.~Potterveld,$^{2}$
R.~D.~Ransome,$^{18}$
J.~Reinhold,$^{4,19}$
E.~Rollinde,$^{19,22}$
O.~Rondon,$^{20}$
P.~Roos,$^{9}$
A.~J.~Sarty,$^{5}$
R.~Sawafta,$^{14}$
E.~C.~Schulte,$^{8}$
E.~Segbefia,$^{7}$
C.~Smith,$^{20}$
S.~Stepanyan,$^{23}$
S.~Strauch,$^{18}$
V.~Tadevosyan,$^{23}$
L.~Tang,$^{7,19}$
R.~Tieulent,$^{9,19}$ 
A.~Uzzle,$^{7}$
W.~F.~Vulcan,$^{19}$
S.~A.~Wood,$^{19}$
F.~Xiong,$^{10}$
L.~Yuan,$^{7}$
M.~Zeier,$^{20}$
B.~Zihlmann,$^{20}$
and V.~Ziskin$^{10}$.
}

\address{
$^{1}${American University, Washington, D.C. 20016} \\
$^{2}${Argonne National Laboratory, Argonne, Illinois 60439} \\
$^{3}${University of Colorado, Boulder, Colorado 80309} \\
$^{4}${Florida International University, University Park, Florida 33199} \\
$^{5}${Florida State University, Tallahassee, Florida 32306} \\
$^{6}${The George Washington University, Washington, D.C. 20052} \\
$^{7}${Hampton University, Hampton, Virginia 23668} \\
$^{8}${University of Illinois, Champaign-Urbana, Illinois 61801} \\
$^{9}${University of Maryland, College Park, Maryland 20742} \\
$^{10}${Massachusetts Institute of Technology, Cambridge, Massachusetts 02139} \\
$^{11}${University of Michigan, Ann Arbor, Michigan 48109} \\
$^{12}${Mississippi State University, Mississippi State, Mississippi 39762} \\
$^{13}${New Mexico State University, Las Cruces, New Mexico 88003} \\
$^{14}${North Carolina A \& T State University, Greensboro, North Carolina 27411} \\
$^{15}${Ohio University, Athens, Ohio 45071} \\ 
$^{16}${Oregon State University, Corvallis, Oregon 97331} \\
$^{17}${Rensselaer Polytechnic Institute, Troy, NY 12180} \\
$^{18}${Rutgers University, New Brunswick, New Jersey 08855} \\
$^{19}${Thomas Jefferson National Accelerator Facility, Newport News, Virginia 23606} \\
$^{20}${University of Virginia, Charlottesville, Virginia 22901} \\
$^{21}${Vrije Universiteit, 1081 HV Amsterdam, The Netherlands} \\ 
$^{22}${College of William and Mary, Williamsburg, Virginia 23187} \\
$^{23}${Yerevan Physics Institute, 375036, Yerevan, Armenia} \\
}

\newpage
\date{\today}

\begin{abstract}
We report on a detailed study of longitudinal strength in the nucleon
resonance region, presenting new results from inclusive electron-proton 
cross sections 
measured at Jefferson Lab Hall C in the four-momentum transfer 
range $0.2 < $Q$^2 < 5.5$ GeV$^2$. The data have been used to accurately 
perform 167 Rosenbluth-type longitudinal / transverse separations.
The precision 
$R = \sigma_L / \sigma_T$ data are presented here, along with the first 
separate values of the 
inelastic structure functions $F_1$ 
and $F_L$ in this regime. The resonance longitudinal component is found 
to be significant, both in magnitude and in the existence of defined
mass peaks. Additionally,
quark-hadron duality 
is here observed above $Q^2 = 1$ GeV$^2$ in the separated structure functions 
independently. 

\end{abstract}


\maketitle



The description of hadrons and their excitations in terms of elementary
quark and gluon constituents continues to be one of the fundamental challenges in physics
today.
Considerable information on nucleon structure has been extracted over the 
past few decades from separations of inclusive lepton-nucleon
cross sections into 
longitudinal and transverse structure functions.  
The original experimental 
observation \cite{bodek} of the vanishing ratio
$R = \sigma_L / \sigma_T$, the ratio of the contributions to the 
measured cross section
from longitudinally and transversely polarized virtual
photon scattering, respectively, as measured 
in deep inelastic scattering (DIS), provided the first evidence of the
fundamental  
spin-1/2 nature of the partons. Since that time, separated 
structure functions have been measured in 
DIS over a wide range of four momentum transfer, $Q^2$, and Bjorken scaling 
variable $x=Q^2/2M\nu$, where $\nu = E - E'$ is the electron energy transfer, 
and M is the proton mass. 

The quantity 
$R$ is expressed in terms of the fundamental nucleon structure functions
$F_1$ (purely transverse), $F_L$ (purely longitudinal), and $F_2$ (combined
longitudinal and transverse) as follows:

\begin{equation}
\label{Rdef}
R \equiv {\sigma_L\over\sigma_T} \equiv { F_L\over{2xF_1} } = 
{ 
{F_2\over{2xF_1}}\left( {1 + {{4M^2x^2}\over{Q^2}} } \right) - 1   
}.
\end{equation}


Precision measurements of $R$ 
are necessary for several fundamental measurements. Extractions 
of the structure function $F_2$, or of the purely 
longitudinal or transverse structure functions, $F_L$ and $F_1$, 
from cross section measurements have historically
depended on assumptions for $R$. The uncertainties introduced 
by this assumption are highly $\epsilon$-dependent, where $\epsilon$ is the
relative longitudinal polarization of the virtual photon in the 
electron-nucleon scattering process. 
Uncertainties in the separation of unpolarized structure 
functions also
have a direct impact on the extraction of the spin structure functions from
spin-asymmetry measurements in electron scattering.
Additionally, precision measurements of $R$ can greatly aid efforts to develop
decisive global descriptions of existing inclusive electroproduction data
at moderate to high $x$ and $Q^2$, necessary for lepton-nucleon scattering model 
development, structure function evolution studies, and accurate radiative 
correction calculations.

Very few measurements of $R$ have been made in the nucleon 
resonance region. Here, the quantity and precision of the existing data (prior
to this work) was such 
that it was impossible to study either the mass-squared ($W^2 = M^2 + 2M\nu - Q^2$) or $Q^2$
dependences of the 
separated longitudinal and transverse resonant structure.
In a resonance excitation 
probed at moderate momentum transfer, 
large values of $R$ and, correspondingly, 
$F_L$, are possible, due to 
gluon exchanges between the quarks.
These effects, as well as the longitudinal character of individual 
resonances, are 
accessible via precision measurements of $R$.
The results presented here represent the first detailed study
of longitudinal strength in the full nucleon resonance region, to  
investigate nucleon resonance structure, and nucleon structure function 
behavior.

The experiment (E94-110) ran in Hall C at the Thomas Jefferson
National Accelerator Facility (Jefferson Lab, or JLab).
An electron beam with a near constant current of 60 $\mu A$ was 
provided by the CEBAF accelerator with seven different 
energies ranging from 1.2 GeV to 5.5 GeV.
Incident electrons were scattered from a 4 cm long liquid hydrogen target and
detected in the High Momentum Spectrometer (HMS), over
an angular range $12.9^{\circ} < \theta < 79.9^{\circ}$. 
To account for backgrounds
from $\pi^0$ production and decay into two photons and subsequent 
electron-positron pairs, positrons were measured 
in the Short Orbit Spectrometer (SOS) and also intermittently in the 
HMS. Other backgrounds included 
electron scattering from the aluminum walls of the
cryogenic target cell, as well as electroproduced negatively charged pions.  
Events from the former were subtracted by performing substitute empty target 
runs, while events from the latter were identified and 
removed by the use of both a gas Cherenkov counter 
and an electromagnetic calorimeter. In all aspects of this experiment, particular
attention was given to demonstrable systematic uncertainty minimization.
For more details regarding the analysis and the Hall
C apparatus employed in this experiment, see Ref. \cite{yl,eric}. 

\begin{figure}
\epsfxsize=3.0in
\epsffile{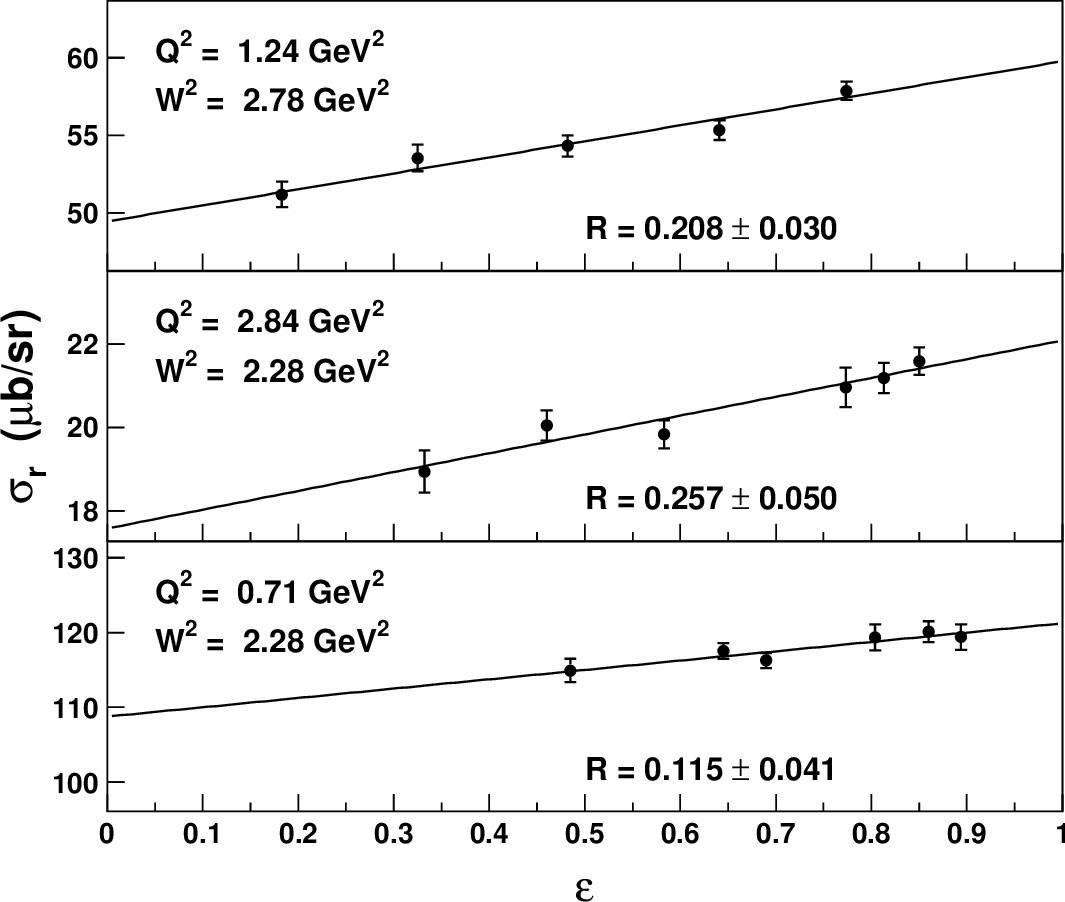}
\caption{Representative Rosenbluth plots for the kinematics indicated.}
\label{Rosen}
\end{figure}

The inclusive double differential cross section for each energy and 
angle bin within the spectrometer acceptance was determined from
  
\begin{equation}
\frac {d \sigma} {d \Omega dE^{'}} = 
\frac{Y_{corr}}{L \Delta \Omega \Delta E^{'}},
\end{equation}

\noindent where 
$\Delta \Omega$ ($\Delta E^{'}$) is the bin width in solid angle 
(scattered energy), $L$ is 
the total integrated luminosity, and  $Y_{corr}$ is the measured
electron yield after correcting for detector 
inefficiencies, background events, and radiative corrections. The latter
include the bremsstrahlung, vertex corrections, and loop diagrams standard
to electron scattering experiments. No corrections were made for 
higher order radiative processes 
involving two photon exchange, since there exists
no decisive inelastic calculation for such effects.
Moreover, minimal (less than $2\%$ \cite{Vladas}) 
non-linear $\epsilon$ dependence of the reduced cross section was observed over
the large 
kinematic coverage in $\epsilon, W^2$, and $Q^2$ of 
this experiment.

For each energy bin, a weighted average cross section 
over $\theta$ within the spectrometer acceptance was obtained
after using a model to correct 
for the angular variation of the cross section from the central angle value.  
In order to minimize dependence on the model used to compute both this and
the radiative correction, 
the following iterative procedure was employed: a model was used to
compute the corrections; the data thus obtained 
were fit to obtain a new model; and this resultant new model was then employed
to recompute the original corrections.
These steps were repeated until the fitting yielded no further
changes. Differing starting models were used to verify 
that the final cross sections were
independent of the starting model within $0.6\%$. A positive byproduct of this 
approach is the availability of a new resonance region fit which 
describes the data here presented to better than $3\%$\cite{eric_fit,webpage1}.

Typical cross section statistical uncertainties 
per energy bin were less than 1$\%$ with systematic errors, 
uncorrelated in $\epsilon$, of $1.6\%$. The total systematic scale
uncertainty in the cross section measurements was $1.9\%$.  
The full cross section sample consisted of 32 scans across the mass-squared 
range  $M^2 < W^2 < 4$ GeV$^2$. Measurements at over 
1,000 distinct $W^2$, $Q^2$ and $\epsilon$ 
points were obtained, allowing for longitudinal / transverse
separations to be performed at $167$ fixed $W^2$, $Q^2$ 
values with typically between 3 and 5 $\epsilon$ points in each separation.

The extractions of purely longitudinal and transverse cross sections and  
structure functions were accomplished via 
the Rosenbluth technique \cite{rosen}, where measurements are made over a 
range in $\epsilon$ at fixed $x$, $Q^2$,
and the reduced cross section, $\sigma_r = d\sigma/\Gamma = \sigma_T + 
\epsilon\sigma_L = \sigma_T(1 + \varepsilon R)$ 
is fit linearly with $\epsilon$. Here, $\Gamma$ is the 
transverse virtual photon flux in the electron-nucleon scattering 
process. Both $\epsilon$ and $\Gamma$ were calculated from 
the measured kinematic variables. The intercept of such a fit gives the 
transverse cross section
$\sigma_T$ (and therefore the structure function $F_1(x,Q^2)$), while the 
slope gives the longitudinal cross section $\sigma_L$, from which can be 
extracted the structure functions $R(x,Q^2)$ and $F_L(x,Q^2)$.  
Because $R$ is determined by the slope of the 
fit, relative to $\sigma_T$, the uncertainty in the extracted value of $R$ 
(and likewise, $F_L$) is dominated by 
the uncorrelated uncertainties in the cross sections versus 
$\epsilon$.  Typical example
Rosenbluth plots are shown Fig. ~\ref{Rosen}.

Prior to a separation being performed, data within a 
$Q^2$ range of $\pm 0.5$ GeV$^2$ and $W^2$ range of $\pm 0.05$ GeV$^2$ for
$W^2 < 3.0$ GeV$^2$ and $\pm 0.10$ GeV$^2$ for
$W^2 \ge 3.0$ GeV$^2$ were brought to a central value using a fit. 
(Larger ranges were employed at the higher W$^2$ values where the cross section becomes 
less W$^2$-dependent.)
Different fits were utilized to assess any model-dependent uncertainty 
in this step, which
was typically less than $3\%$. This uncertainty concern dictated that
separations were not performed if the required centering correction was larger
than $60\%$.


\begin{figure}
\epsfxsize=3.0in
\epsffile{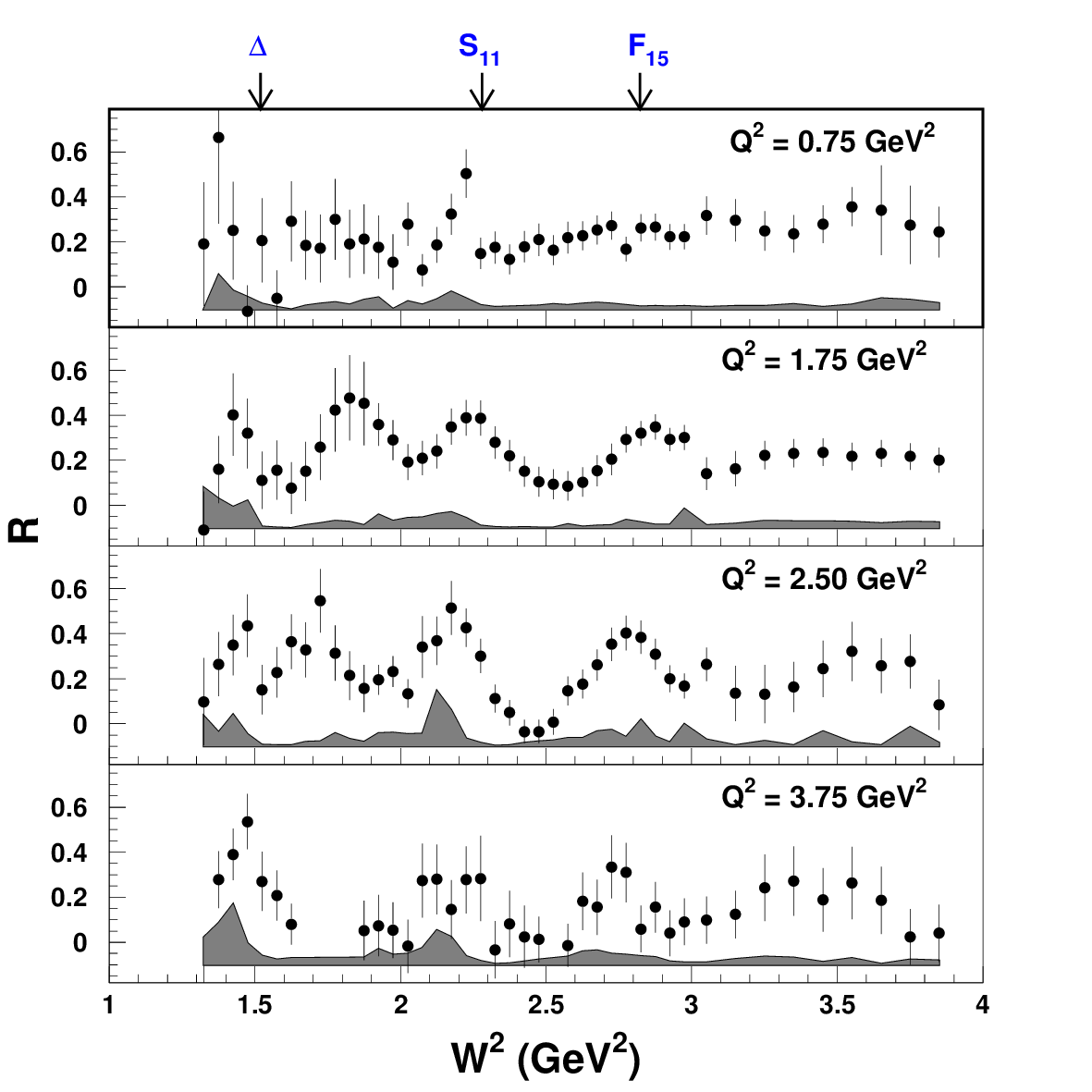}
\caption{Measurements of $R=\sigma_L/\sigma_T$, as a function of $W^2$, for
the $Q^2$ values indicated. The error bars shown represent both the 
statistical and uncorrelated systematic uncertainties, with 
the former negligible in comparison to the latter. The shaded band 
below the offset zero represents the 
total scale uncertainty. The locations of the three prominent resonances observed in the
unseparated cross section measurements are labeled at the top.}
\label{rvsw2}
\end{figure}

Values obtained for $R$ are plotted versus $W^2$ in 
Fig. ~\ref{rvsw2} for the $Q^2$ values indicated. 
It is clear from the mass enhancements in the data that $R$ exhibits 
resonant structure, and that 
this variation with $W^2$ can be quite large. This is the first direct 
observation of such structure, contradicting a common assumption
that the resonance
contribution to $R$, or the longitudinal strength in the resonance region,
is small or negligible (for example, \cite{paul,linda,dalton,k2,k3,k4,k5}). 

The almost twenty
well-established nucleon resonances with masses below $2$ GeV
give rise to only three distinct enhancements in the unseparated
inclusive electron scattering cross
section and, of the three, only the first (the lowest mass 
$P_{33}(1232)$, $\Delta$) state
is not a superposition of overlapping resonant states. 
The second enhancement region is often
referred to as the $S_{11}$, as the unseparated 
cross section here is dominated above $Q^2 \approx 1$ GeV$^2$ by the ground 
state $S_{11}(1535)$ resonance, 
even though there exists the $D_{13}(1520)$ overlapping resonance. 
A similar situation is true
for the third enhancement region, which is dominated by the $F_{15}(1680)$, 
with even more overlapping resonance states. 
In $R$, however, the situation appears to be different. There is an additional
prominent peak at $W^2 \approx 1.8$ GeV$^2$, somewhat below the 
$S_{11}$ dominated mass region. 

The lowest mass $\Delta$ resonance region exhibits non-negligible longitudinal 
strength, which does not appear to 
diminish over the $Q^2$ range of this experiment. The spin-flip 
required for this positive parity isospin $I = {3\over2}$ 
excitation suggests that it be dominantly transverse, 
yet some models 
predict a possible non-negligible longitudinal component \cite{Kroll,kl5,kl6,kl7,kl8}. 
Recent analyses of predominantly exclusive scattering 
data from JLab \cite{HallBlong,Kelly_long,mesonrev} differ somewhat but generally tend to 
indicate a small longitudinal $\Delta$
resonant component. It has been noted that Rosenbluth separated 
data such as that presented here will be critically useful input to such
analyses \cite{Kelly_short}. It is also possible
that the $R$ values here observed may indicate a substantial non-resonant
background contribution in this regime. 

\begin{figure} [tb]
\epsfxsize=3in
\epsffile{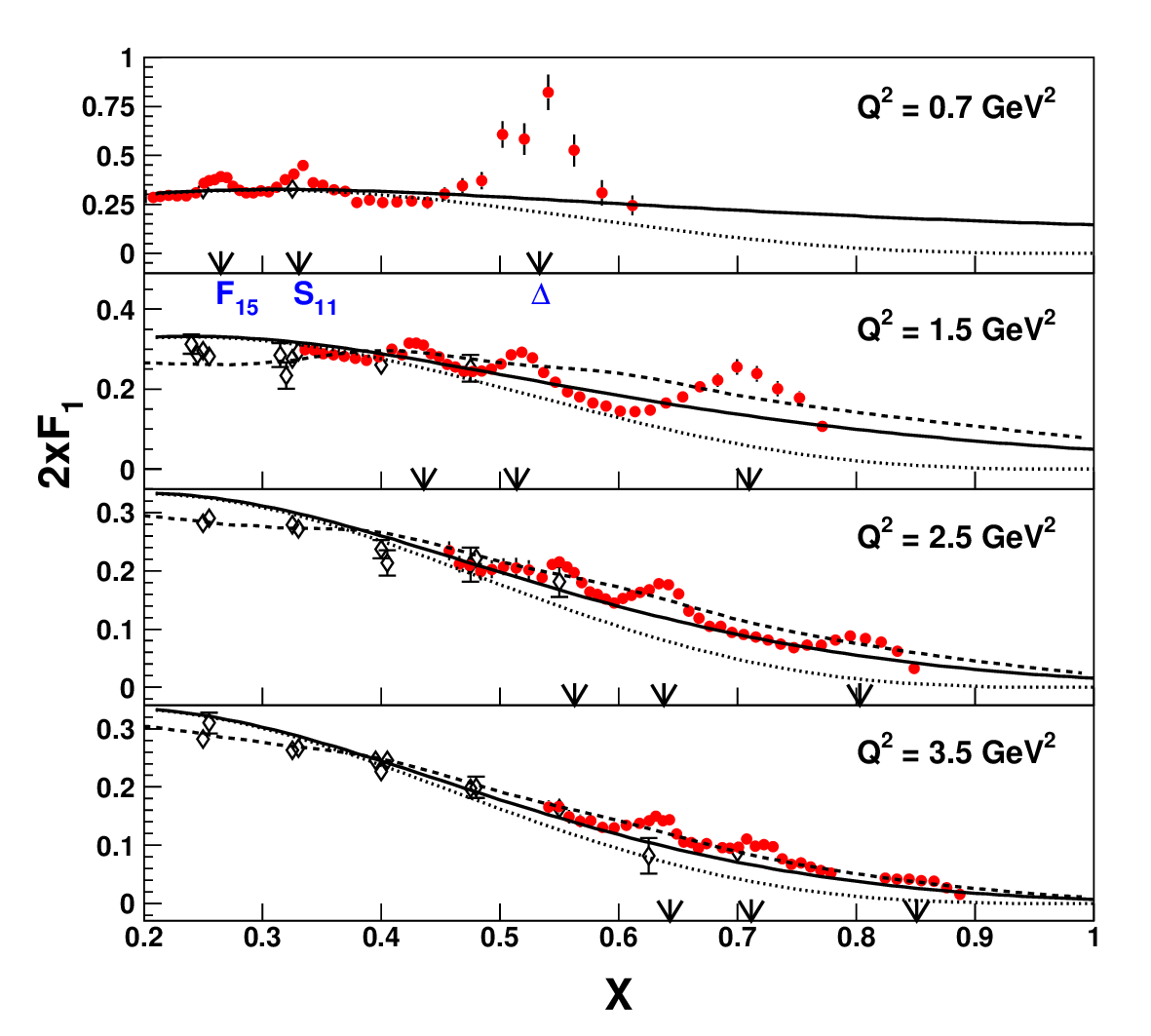}
\epsfxsize=3in
\epsffile{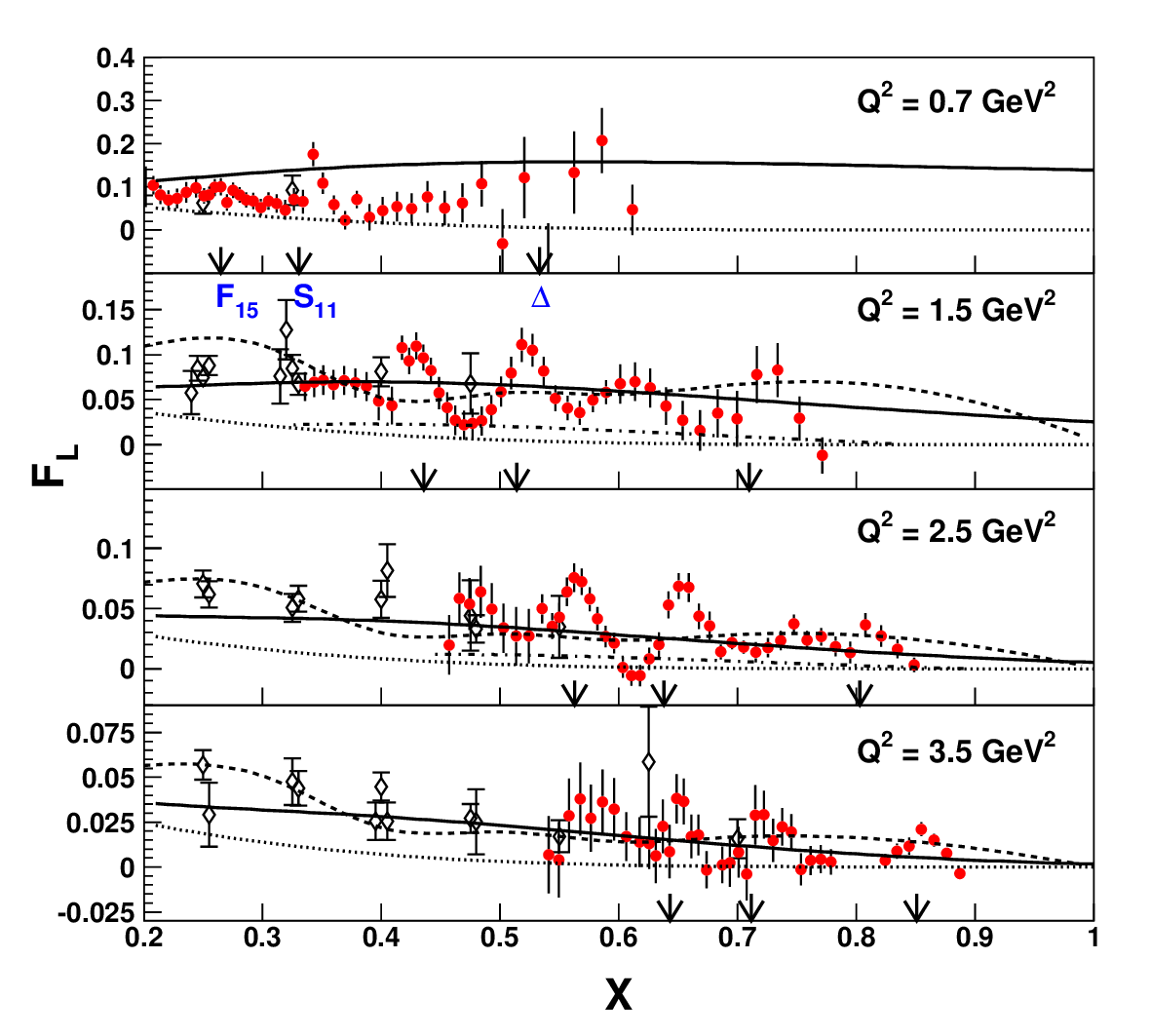}
\caption{The longitudinal structure function $F_L$ (top), 
and transverse nucleon structure function $2xF_1$ (bottom), measured 
in the resonance region (triangles) as a function of $x$, compared 
with existing DIS measurements from SLAC (diamonds). The curves
are from MAID (bottom middle, dot-dashed), Alekhin (dashed), 
and MRST with (solid) and 
without (dotted) target 
mass effects included. The three prominent 
resonance mass regions observed in the inclusive cross section 
are indicated by arrows, and labeled in the top plots. The error bars shown 
represent both the statistical and systematic uncertainties, 
with the latter being dominant.}
\label{f1fl}
\end{figure}

The possible peak observed in $R$ at $W^2 \approx 1.8$ GeV$^2$ 
in Fig. ~\ref{rvsw2} below the $S_{11}$ is notable, albeit with large
uncertainty, in the $F_L$
longitudinal channel. This mass is close to that of the elusive Roper
resonance, $P_{11}(1440)$, the electroproduction of 
which is a topic of some interest (see for example, 
\cite{HallBroper,Lee,ARuso,Volker,Beane,Matheus,Jaffe}). The excitation of the 
Roper resonance has been found to be dominantly longitudinal \cite{Kelly_long}.
The observed mass is also
near the $P_{13}$, or $\Sigma(1385)$, resonance. This resonance should have a 
small cross section in electroproduction, however it could show up
preferentially in the longitudinal channel which is dominant in kaon 
production. Regardless of its origin, this is a surprising observation of 
significant resonant longitudinal strength, and bears further experimental and
theoretical investigation. 


In all, the data clearly exhibit differing longitudinal and transverse resonance 
behavior, as shown in Fig. ~\ref{f1fl}, where the purely transverse $2xF_1$ and
purely longitudinal $F_L$ structure functions are plotted separately.
Here, the structure functions are plotted as a function of $x$ rather than
$W$, for the purpose of further discussions below. The mass peak
regions move up in $x$ with
increasing $Q^2$. It may be observed that there are mass peaks in {\it both} the
longitudinal and transverse channels, and that the peak positions differ somewhat. 
Not only do the data unequivocally demonstrate 
significant longitudinal resonance structure, but the $W$-dependence of 
$F_L$ is larger than that of $2xF_1$ above $Q^2 \approx 1$ GeV$^2$, as evidenced
by the relatively greater prominence of the mass peaks. 

A precise extraction of information on individual resonances, 
such as transition form factors, from this inclusive 
data must involve a detailed fitting study beyond the scope of this report.
At lower values of $Q^2 < 1$ GeV$^2$, unitary isobar fits like MAID 
\cite{MAID} give quite definite and accurate predictions based on 
single pion, two pion, 
eta, and kaon decay channels for resonances below $W^2 < 4$ GeV$^2$. 
At the higher $Q^2$ values of the data presented here, however, multi-pion 
effects, tails of higher mass resonances, and non-resonant components are
very significant and therefore such fits tend to underestimate this data, as can be 
seen from the MAID curve in Fig. ~\ref{f1fl} (bottom). 

Also presented with the resonance region data 
in Fig. ~\ref{f1fl} are the predominantly DIS ($W^2 > 4$ GeV$^2$) data 
from Rosenbluth separations 
performed at SLAC \cite{dasu,tao}. Where overlapping, 
the two data sets are in 
agreement, providing additional confidence in the achievement of 
the demanding precision required for these experiments. 
In all cases, there is a smooth transition between the resonance and DIS data 
in both $x$ and $Q^2$. 

The curves shown are parton distribution based parameterizations of structure
functions at
next-to-next-to leading order, from Alekhin \cite{Alekhin}, including 
target mass effects according to \cite{GP}, and from
MRST \cite{MRST}, both with and 
without target mass effects according to \cite{Barbieri,TMreview}. 
The MRST parameterization includes data from deep inelastic scattering
as well as other experiment types, while Alekhin's calculation uses only DIS. 
The latter calculation is valid only down to $Q^2 = 1$ GeV$^2$.

It is clear that some prescription for 
target mass effects is required to describe the data. 
However, for $Q^2 > 1$ GeV$^2$, it appears that 
minimal if any additional non-perturbative descriptions (such as higher twist effects)
seem necessary to describe 
the {\it average} behavior of the resonance region. The resonances
oscillate around the scaling curves. Furthermore, this
is true for the range of different $Q^2$ values, indicating that the scaling
curve describes as well the average $Q^2$ dependence of the resonance
regime. These observations are consistent with 
quark-hadron duality \cite{dual_review}, and may
be counted as the first observation of duality in the separated
transverse and longitudinal structure functions. 

In summary, we have reported results from 
a first detailed study of longitudinal and transverse strength 
in the nucleon resonance region. The new data have yielded an array of
interesting observations. Contrary to most transition 
form factor fit assumptions, the resonant
longitudinal component is found to be 
substantial. Furthermore, the resonance mass dependence of the longitudinal 
structure function is more pronounced than the transverse. 
Significant strength is observed between the $S_{11}$ 
and $P_{33}$ resonance mass regions in the longitudinal channel. 
Separated 
measurements of the inelastic structure functions $F_1$ 
and $F_L$ are presented. The data show quark-hadron duality for the first time
in the $F_1$ and $F_L$ structure functions independently.
 
These data are now available \cite{webpage2} for additional studies.


We gratefully acknowledge research grant support from the  
National Science Foundation and the U.S. Department of Energy. 
We thank the Jefferson Lab Hall C scientific
and engineering staff for their outstanding support. The Southeastern
Universities Research Association operates the Thomas Jefferson National
Accelerator Facility under the U.S. Department of Energy contract 
DEAC05-84ER40150.


\begin{thebibliography}{99}
\bibitem{webpage2} The data are available at hallcweb.jlab.org/resdata.
\end{thebibliography}
\end{document}